\documentstyle[12pt]{article}
\def\abstract#1{{\centerline{\bg Abstract}} \vskip 3mm \par #1}
\def\cy{Calabi-Yau}
\def\cym{Calabi-Yau manifold}
\def\lg{Landau-Ginzburg}
\def\lgo{LG orbifold}
\def\p{Poincar\'{e}}
\def\pp{Poincar\'{e} polynomial}
\def\pps{Poincar\'{e} polynomials}
\def\tt{\tilde{\theta_{i}}}
\def\tg{\tilde{\theta_{i}}^{g}}
\def\th{\tilde{\theta_{i}}^{h}}

\def\ep{\epsilon}
\def\inbar{\vrule height1.5ex width.4pt depth0pt} 
\def\ZZ{\relax{\sf Z\kern-.4em \sf Z}}  \def\IR{\relax{\rm I\kern-.18em R}}
\def\IN{\relax{\rm I\kern-.18em N}} \def\IP{\relax{\rm I\kern-.18em P}}
\def\IQ{\relax\,\hbox{$\inbar\kern-.3em{\rm Q}$}}
\def\la{\lambda}
\def\IC{\hbox{\,$\inbar\kern-.3em{\rm C}$}}
\begin{document}
\baselineskip=6mm
\begin{flushright}
{KOBE-TH-93-10} \\
{November 1993}
\end{flushright}
\vskip 2.5cm
\centerline{\LARGE On the Poincar\'{e} Polynomials for }
\vskip 1cm
\centerline{\LARGE Landau-Ginzburg Orbifolds}
\vfill
\centerline{\large Hitoshi Sato}
\vskip 1cm
\centerline{\it Graduate School of Science and Technology, Kobe University}
\centerline{\it Rokkodai, Nada, Kobe 657, Japan}
\centerline{email address : UTOSA@JPNYITP.BITNET}
\vfill
\centerline{Abstract}
\par
We construct the \pp s for Landau-Ginzburg orbifolds with projection
operators.
Using them we show that special types of dualities including \p\ duality
are
realized under certain conditions. When Calabi-Yau interpretation exists,
two
 simple formulae for Hodge numbers $h^{2,1}$ and $h^{1,1}$ are obtained.
\vfill
\thispagestyle{empty}
\newpage
\pagenumbering{arabic}
\section{Introduction}

A compactified heterotic string theory with $N=1$ space-time supersymmetry
requires an $N=2$ superconformal field theory (SCFT) for its internal
space.
Although at least (2,0) world sheet supersymmetry is necessary to have
$N=1$
space-time supersymmetry, models with (2,2) supersymmetry are of special
interest because they could have geometrical picture. In fact, a large
class
of such models with central charge $c=9$ allows a \cy\ interpretation
\cite{g}.

It has been observed that \lg\ models can uncover the relation between
$N=2$
SCFT and \cy\ picture \cite{gvw}. A \lg\ model describes a chiral ring of
 $N=2$ SCFT. It is known that a large class of \lg\ models with integral
 $U(1)$ charge states have \cy\ interpretation by identifying a chiral
primary
  state which has $U(1)$ charge ({\it\^{c}-p,q}) with a ({\it p,q}) form on
   a \cym, where $\hat{c}=c/3$. If \cy\ interpretation exists, the number
   of non-singlet representations of $E_{6}$ can be obtained from the
charge
    degeneracies of a \lg\ model.

To get low generation models, Landau-Ginzburg (LG) orbifold models and
corresponding \cy\ orbifolds have recentry been considered. \lgo s were
 first considered by Vafa \cite{v} and further results are obtained in
 ref.\cite{iv}. Under certain conditions \lgo s still have \cy\
interpretation
 and many examples are found \cite{ks2,ks3,kss}. These models reveal a
special
 feature called mirror symmetry \cite{cls,gp,bh}, which is expected to give

 an effective method for calculating Yukawa couplings \cite{c,cop}.
same 

In this paper we investigate the structures of \lgo s. To do this, \pp s
for
 \lgo s with projection operators are good tools. However, such \pp s are
not
  known except for the models divided by $U(1)$ projection operator $j$
only.
   So we first construct \pps\ with general projection operators for
certain
    calsses of \lgo s.
 In geometrical picture there are some dualities of Hodge numbers on a
\cym.
In the case of \lgo s we can find, through calculations model by model,
that
there exist dualities which are equivalent to those of \cy\ orbifolds.
But it is not clear why they are realized.
Using our \pp\ we show how special dualities of \lgo s are realized.
Finally we obtain simple fomulae for dimensions of chiral prmary fields.
 These dimensions correspond to undetermined Hodge numbers $h^{1,1}$ and
  $h^{2,1}$, so they determine the generation number of a heterotic string
  model compactified on a \cym\ \cite{chsw}.

This paper is organized as follows; in section 2 we construct the \pp s for
 \lgo s with projection operators. It is shown in section 3 that these
polynomials
  satisfy special types of dualities. Two fomulae for dimensions of chiral
primary
   fields are obtained in section 4. We give conclusions and discussions in
the
   last section.
\section{Construction of the \pp}

The Landau-Ginzburg description of chiral rings depends on the assumption
that
 an N=2 supersymmetric theory with a non-degenerate quasi-homogeneous
 superpotential $W(\la^{n_{i}} X_{i}) = \la^{d}W(X_{i})$, $i=1 \sim N$,
flows
 to an N=2 superconformal model at a renormalization group fixed point. The

 charges of the chiral super fields $X_{i}$ are $q_{i} = n_{i}/d$ and the
central
 charge is given by $c = 3\sum_{i}{(1-2q_{i})}$. The chiral ring ${\cal R}$
is
 a quotient ring defined by

\begin{equation}
\label{cr}
{\cal R} = \frac {\IC [X_{i}]}{[\partial_{j} W]},
\end{equation}
and the \pp\ for ${\cal R}$ is given by \cite{lvw}

\begin{equation}
{\bf Tr}_{(c,c)}[t^{J_{0}}\bar{t}^{\bar{J}_{0}}]=\prod_{i}
{\frac{1-(t\bar{t})^{(1-q_{i})}}{1-(t\bar{t})^{q_{i}}}}
\end{equation}

The \lgo s are obtained by quotienting with an abelian symmetry group $G$
of $W(X_{i})$,  whose element $g$ acts as an $N \times N$ diagonal matrix;

\begin{equation}
g: X_{i} \rightarrow e^{2 \pi i {\theta_{i}}^{g}}X_{i}.
\end{equation}
Only the $X_{i}$ invariant under the above transformation, that is those
for which $\tg \equiv {\theta_{i}}^{g}-[{\theta_{i}}^{g}]$ vanishes
contribute
 to the chiral ring in the $g$-twisted sector. For a geometrical
interpretation
  the important thing is $U(1)$ projection operator $j$, which acts as
$j: X_{i} \rightarrow e^{2 \pi i {q_{i}}}X_{i}$.

\pp s for such \lgo s have already been constructed by Intriligator and
Vafa
 in ref.\cite{v,iv}. But they do not contain informations about
transformation
  proprties of the states under $g$, so that the states which shoud be
projected
   out still remain. So it is necessary to construct the \pp\ for a \lgo\
with
    a projection operator in order to know the true chiral ring structure.

We will restrict ourselves to the following three types of polynomials or
 products of them for $W(X_{i})$;
\begin{enumerate}
\item $X_1^{a_1}+ \cdots +X_n^{a_n}$
\item $X_1^{a_1}X_2+X_2^{a_2}X_3+ \cdots +X_n^{a_n}X_1$
\item $X_1^{a_1}X_2+X_2^{a_2}X_3+ \cdots +X_n^{a_n}$.
\end{enumerate}
They are called Fermat, loop and tadpole type, respectively. These three
types
of $W(X_{i})$ are very interesting because they contain A-D-E $N=2$\
supersymmetric
minimal models. Note that only these types appear in the
Berglund-H\"{u}bsch
 construction of mirror manifolds\cite{bh}.

In order to obtain \pp\ with $g$ we first construct the chiral ring
(\ref{cr})
 directly.
Before explaining our method, let us briefly summarize the results in
\cite{v,iv}.
The arbitrary (c,c) states in the $h$-twisted sector can be represented as
\begin{equation}
\prod_{\th=0} {(X_{i})^{\ell_i}}{\Big|0\Big\rangle^{(h)}_{(c,c)}}
\end{equation}
with appropriate $\ell_{i}$.
The transformation properties of chiral primary states have been
investigated
from modular invariance of the Witten index and they are given by

\begin{equation}
\label{tcc}
g \prod_{\th=0} {(X_i)^{\ell_i}}{\Big|0\Big\rangle^{(h)}_{(c,c)}} =
e^{2 \pi i \phi^{g}(h)} (\exp 2 \pi i\sum_{\th = 0}{\tg\ell_{i}})
\prod_{\th=0}
{(X_i)^{\ell_i}}{\Big|0\Big\rangle^{(h)}_{(c,c)}},
\end{equation}
where
\begin{eqnarray*}
e^{2 \pi i \phi^{g}(h)}=\ep(g,h)(-1)^{K_{g}K_{h}}(\det g)^{-1}\det g
\mid_{h}
  \end{eqnarray*}
and
\begin{eqnarray*}
\det g \mid_{h} = \exp (2 \pi i\sum_{\th = 0}{\tg}).
\end{eqnarray*}

$\ep(g,h)$ is the discrete torsion and $K_g$ is an integer defined mod 2.
 They are not determined by modular invariance of the Witten index. But
they
 must satisfy certain conditions to get integral left and right charges. We
will
  demand such conditions in later sections.

The charges of the state
${\Big|0\Big\rangle^{(h)}_{(c,c)}}$
were obtained \cite{v} to be
\begin{equation}
\label{uc}
\begin{array}{cc}
\left(\begin{array}{c}
J_{0} \\
\bar{J_{0}}
\end{array} \right) & {\Big|0\Big\rangle^{(h)}_{(c,c)}} = (\sum_{\th>0}
{(\frac{1}{2}-q_{i}) \pm
(\th-\frac{1}{2})}){\Big|0\Big\rangle^{(h)}_{(c,c)}}.
\end{array}
\end{equation}

Now we turn our attention to the \pp. To illustrate our construction we
consider a Fermat type as an example.
For a Fermat type polynomial, the chiral ring ${\cal R}$ is determined by
the equations

\begin{equation}
{X_{i}}^{a_{i}-1} = 0.
\end{equation}
and chiral primary states are in the form $\prod_{i}{X_{i}^{{\ell}_{i}}}$
with $0 \leq \ell_{i} \leq {a_{i}-2}$.
To construct a \pp\ it is useful to introduce the notations $x_{i} \equiv
 t^{q_{i}}$. In this case it is easy to see that the \pp\ is given by

\begin{equation}
P(t) = \frac{{x_{1}}^{a_{1}-1}-1}{x_{1}-1}
\frac{{x_{2}}^{a_{2}-1}-1}{x_{2}-1}
\cdots \frac{{x_{n}}^{a_{n}-1}-1}{x_{n}-1}.
\end{equation}

Twisted sectors appear when the \lg\ theory is orbifoldized. In the
$h$-twisted
sector we must consider the following restricted polynomial
\begin{eqnarray*}
\prod_{\th = 0}{\frac{{x_{i}}^{a_{i}-1}-1}{x_{i}-1}}.
\end{eqnarray*}

When $g$ acts on this chiral ring, the \pp\ gets phases according to
eq.(\ref{tcc}).
 These phases are obtained by transformations

\begin{equation}
x_{i} \rightarrow e^{2 \pi i \tg}x_{i}.
\end{equation}
except for the overall phase $e^{2 \pi i \phi^{g}(h)}$, which is defined
 below (\ref{tcc}).

Then we obtain the \pp\ for $(c,c)$ states of the $h$-twisted sector,

\begin{equation}
\label{pp}
{\bf Tr}_{(c,c)}[gt^{J_{0}}\bar{t}^{\bar{J}_{0}}] \mid_{h-twisted}=
e^{2 \pi i \phi^{g}(h)}\prod_{\th=0}\frac{1-e^{2 \pi i (1-\tg)}
(t\bar{t})^{(1-q_{i})}}{1-e^{2 \pi i \tg}(t\bar{t})^{q_{i}}}\prod_{\th>0}
(t\bar{t})^{(\frac{1}{2}-q_{i})}(t/\bar{t})^{(\th-\frac{1}{2})},
\end{equation}
where we have changed the variables from $x_{i}$ to $t^{q_{i}}$ and the
trace
 has been taken over the $h$-twisted sector only.
The last product arises due to (\ref{uc}).
For loop and tadpole type polynomials we can get the same results
\cite{sa}.

Consequently we can construct the \pp\ with the projection operator
 $\hat{P}\equiv\frac{1}{\mid G \mid} \sum (all\  g \in G)$;

\begin{equation}
\label{ppp}
 P(t,\bar{t}) \equiv {\bf
Tr}_{(c,c)}[\hat{P}t^{J_{0}}\bar{t}^{\bar{J}_{0}}],
\end{equation}
where we have taken the trace over all the sectors. This \pp\ contains
informations on projected (c,c) states only. The coefficients of the \pp\
$p_{i,j}$
 defined by $P(t,\bar{t}) = \sum p_{i,j} t^{i}\bar{t}^{j}$ are the
dimensions of
  the chiral primary fields with charge $(i,j)$.
\section{Special dualities}
It is well known that Hodge numbers of a \cy\ 3-fold obey the following
relations,
 which we call special dualities \cite{h};

\begin{enumerate}
\item \p\ duality; $h^{i,j}=h^{3-i,3-j}$
\item Complex conjugation duality; $h^{i,j}=h^{j,i}$
\item Holomorphic duality; $h^{0,j}=h^{0,3-j},\ h^{i,0}=h^{3-i,0}$.
\end{enumerate}

We will show that for $p_{i,j}$ the same special dualities hold if an \lgo\
has
a geometrical interpretation. Due to the identifications $p_{\hat{c}-i,j} =
 h^{i,j}$, these dualities are 1. $p_{i,j} = p_{\hat{c}-i,\hat{c}-j}$ ;
  2. $p_{\hat{c}-i,j} = p_{\hat{c}-j,i}$ and 3. $p_{\hat{c},j} =
p_{\hat{c},3-j},
  \ p_{\hat{c}-i,0} = p_{i,0}$. In the following we set $c = 9$. To see
these
  dualities we should know which twisted sector contributes to $p_{i,j}$.
  We first consider $p_{i,0}$ and $p_{0,i}$ $(i=1,2)$, because some \lgo s
  could have non-vanishing ones.

{}From our \pp\ (\ref{pp}) nonzero $p_{2,0}$ can arise only from the twisted
sector with both $\sum_{\th > 0}{(\th - q_{i})} =2$ and $\sum_{\th > 0}
{(1 - \th - q_{i})} =0$.
Using the similar discussion by Kreuzer and Skarke \cite{ks2}, we can show
 that $\sum_{\th > 0}{(1 - \th - q_{i})} =0$ implies $(1 - \th -q_{i}) = 0$

 for $\th > 0$ \cite{sa}, so that in this case we have $\sum_{\th > 0}{(1 -
2q_{i})}
  = 2$. We therefore conclude that nonzero $p_{2,0}$ can exist only
  if the following two conditions are satisfied;
\begin{enumerate}
\item There is at least one element $j_{2}$ of the symmetry group $G$ such
that $j_{2}$ acts like $j$ on the fields $X_{i} \in {\cal M}_{2}$
(where ${\cal M}_{2}$ is a subset of the fields {$X_{i}$}) and does not
act at all on the other fields.
\item The contribution of the fields in ${\cal M}_{2}$ to the central
charge
is 6, i.e. $\sum_{X_{i} \in {\cal M}_{2}} (1-2{q_{i}})=2$.
\end{enumerate}
Thus only the $j_{2}^{-1}$-twisted sector can contribute to $p_{2,0}$ and,
 by the above definition, only the $j^{-1}j_{2}$-twisted sector can
contribute
 to $p_{1,0}$.

We can see that the operator $jj_{2}^{-1}$ is nothing but the $j_{1}$ in
 ref.\cite{ks3}. It was shown that nonzero $p_{0,1}$ and $p_{0,2}$ can
exist
 only if there is at least one element $j_{1}$ of $G$ \cite{ks3}. So we see

 that $p_{1,0},\ p_{2,0},\  p_{0,1}$ and $p_{0,2}$ can be nonzero only if
the
 element $j_{1}$ of $G$ exists.

The above discussion allows us to find the twisted sectors which contribute

to other $p_{i,j}$.
In the $h$-twisted sector we can see through the \pp\ (\ref{pp}) that there

is one state with the lowest charges
($\sum_{\th > 0}{(\th-q_{i})}$ , $\sum_{\th > 0}{(1-\th-q_{i})}$)
and that only one state with the highest charges
($Q+\sum_{\th > 0}{(\th-q_{i})}$ , $Q+\sum_{\th > 0}{(1-\th-q_{i})}$)
exists,
 where $Q = \sum_{\th =0}{(1-2q_{i})}$. Then we obtain the results listed
in Table 1.
\begin{table}[h]
\[ \begin{tabular}{||c|c||} \hline
$p_{i,j}$ & twisted sector \\ \hline \hline
$p_{0,0}$ , $p_{3,3}$ & untwisted \\ \hline
$p_{0,3}$ & $j$ \\ \hline
$p_{3,0}$ & $j^{-1}$ \\ \hline
$p_{0,1} , p_{2,3}$ & $j_{1}$ \\ \hline
$p_{1,0} , p_{3,2}$ & ${j_{1}}^{-1}$ \\ \hline
$p_{0,2} , p_{1,3}$ & $jj_{1}^{-1}$ \\ \hline
$p_{2,0} , p_{3,1}$ & ${(jj_{1}^{-1})}^{-1}$ \\ \hline
$p_{2,1}$ & $h_{1}\ \cdots$  \\ \hline
$p_{1,2}$ & ${h_{1}}^{-1}\ \cdots$ \\ \hline
$p_{1,1} , p_{2,2}$ & $h_{2}\ \cdots$ \\ \hline
\end{tabular} \]
\caption{The twisted sectors contributing to $p_{i,j}$.
$h_{1},\ \cdots$ ($h_{2},\ \cdots$) indicate the twisted sectors that
 contribute to $p_{2,1}$ ($p_{1,1}$ and $p_{2,2}$).}
\label{t1}
\end{table}

It was shown in \cite{iv} that, to get an LG orbifold with integral left
and
right charges, we must take $G$ containing $j$ and require $\ep(j,g)=
(-1)^{K_{j}K_{g}}\det g$ and $(-1)^{K_{g}}=\det g$. For geometrical
interpretation we further require $\det g =1$,\ $\ep(g,h)=(-1)^{K_{g}}=1$\
for all
$g$ and $h$ in $G$. Hence we have $\det j=1$ and this implies the number of

fields $N=odd$ for $c=9$. The requirement $\det g = 1$ for all $g \in G$ is
 needed to guarantee the invariance of holomorphic three form \cite{c}.
In the remaining part of this section we assume these requirements to hold.

We can see through the \pp\ (\ref{pp}) that in the $h$-twisted sector
the phase of the state with the lowest charges
is $\det g |_{h}$ and the one with the highest charges
is ${(\det g |_{h})}^{-1}$.
Then we find that the (c,c) states have the phases listed in Table 2,
if they exist.
\begin{table}
\[ \begin{tabular}{||c|c||} \hline
state & phase \\ \hline
(0,1) & $\det g|_{j_{1}} = \prod_{\tt^{j_{1}}=0}{e^{2 \pi i \th}}$ \\
\hline
(2,3) & ${(\det g|_{j_{1}})}^{-1} = \prod_{\tt^{j_{1}}=0}{e^{-2 \pi i
\th}}$
 \\ \hline
(0,2) & ${\det g|_{j{j_{1}}^{-1}}} = \prod_{\tt^{j{j_{1}}^{-1}}=0}{e^{2 \pi
i \th}}$
\\ \hline
(1,3) & ${(\det g|_{j{j_{1}}^{-1}})}^{-1} = \prod_{\tt^{{jj_{1}}^{-1}}=0}
{e^{-2 \pi i \th}}$ \\ \hline
(2,0) & ${\det g|_{{(j{j_{1}}^{-1})}^{-1}}} =
\prod_{\tt^{{(j{j_{1}}^{-1})}^{-1}}=0}
{e^{2 \pi i \th}}$ \\ \hline
(3,1) & ${(\det g|_{{(j{j_{1}}^{-1})}^{-1}})}^{-1} = \prod_
{\tt^{{(j{j_{1}}^{-1})}^{-1}}=0}{e^{-2 \pi i \th}}$ \\ \hline
(1,0) & $\det g|_{{j_{1}}^{-1}} = \prod_{\tt^{j_{1}^{-1}}=0}{e^{2 \pi i
\th}}$
\\ \hline
(3,2) & ${(\det g|_{{j_{1}}^{-1}})}^{-1} = \prod_{\tt^{{j_{1}}^{-1}}=0}
{e^{-2 \pi i \th}}$ \\ \hline
\end{tabular} \]
\caption{The phases of the states with charges $(i,j)$.}
\label{t2}
\end{table}

By definition $\det g|_{j_{1}} = \det g|_{{j_{1}}^{-1}}$, and owing to
 $\det g=1$ for all $g \in G$ we obtain $\det g|_{j_{1}} =
 {(\det g|_{j{j_{1}}^{-1}})}^{-1}$. Therefore we see that if one of the
above
 states survives after projection then all the other states must do. So the
set
  of $p_{i,j}$ with the above charges satisfies special dualities. It can
   be shown through the observation of the phases in eq.(\ref{pp}) that
    if (2,1) states in the $h_{1}$-twisted sector survive then
corresponding
    (1,2) states in the ${h_{1}}^{-1}$-twisted sector must survive, or vice
     versa. We can see that there are the same structures between (1,1) and

     (2,2) states. Therefore each set of \{ $p_{2,1}$ , $p_{1,2}$ \} and
     \{ $p_{1,1}$ , $p_{2,1}$ \} respects the dualites 1. and 2.
Consequently
     we have special dualities for the \lgo\ corresponding to \p\ duality,
     complex conjugation duality and holomorphic duality.

\section{Formulae for $p_{1,1}$ and $p_{2,1}$}
In this section we consider the $p_{1,1}$ and $p_{2,1}$ through the \pp\
(\ref{ppp}). Let $n_{27}$ and $n_{\bar{27}}$ be the number of {\bf 27} and
${\bf \bar{27}}$ representations of $E_6$ in a heterotic string model. If
we have a \cy\ interpretation, then $p_{1,1}=n_{27}$ and
$p_{2,1}=n_{\bar{27}}$.
 So they determine the generation number to be $|p_{2,1}-p_{1,1}|$
\cite{chsw}.

For an \lgo\ with integral left and right charges we have using the \p\
polynomial
 (\ref{ppp}),
\begin{equation}
P(-1,-1)=\ {\frac{1}{ \mid G \mid } \sum_{i}(-1)^{N+K_{g}K_{h}+K_{gh}}
\ep(g,h)\prod_{\tilde{\theta}_i^g=\tilde{\theta}_i^h=0}(1-
{\frac{1}{q_i}})}\ ,
\end{equation}
where $K_{gh}=K_{g}+K_{h}$.
In the following we consider \lgo s with \cy\ interpretations. So we put
the
 same conditions in section 3 and we have $P(-1,-1)=2(p_{1,1}-p_{2,1})$.
The
 identifications $p_{1,1}=h^{2,1}$ and  $p_{2,1}=h^{1,1}$\  imply
$P(-1,-1)=-\chi$.

In ref.\cite{ks3} it was shown that the exponent of $t$ in a \pp\ is
non-negative,
\begin{equation}
\sum_{\th>0}{(\th-q_{i})} \geq 0.
\end{equation}
By a similar discussion we can also show that the exponent of ${\bar t}$ is
 non-negative \cite{sa},
\begin{equation}
\sum_{\th>0}{(1-\th-q_{i})} \geq 0.
\end{equation}
{}From these inequalities we obtain
\begin{equation}
-\frac{\hat{c}}{2}\leq\sum_{\th>0}(\th-\frac{1}{2})\leq \frac{\hat{c}}{2}.
\end{equation}
Let us denote by $I$ the number of invariant fields in the $h$-twisted
sector.
If $I = odd$ , $\sum_{\th>0}(\th-\frac{1}{2})=0,\pm1$.

We consider LG orbifolds with
$\sum_{\tt>0}(\tt-\frac{1}{2})\neq\pm1$
in all the twisted sectors (namely LG orbifolds without $j_{1}$) since the
generation number of the \lgo s with nonzero $p_{0,1}$ vanishes
\cite{ks2,ks3}.
We see that chiral primary states in the sector with $I=odd$ contribute to
 $p_{i,i}$ for $i=0\sim3$ and in the sector with $I=even$ contribute to
other
 $p_{i,j}$. Using the above discussions and structures of $p_{i,j}$, we
have

\begin{equation}
\label{p11}
p_{1,1}= \{ {\frac{-1}{2} \frac{1}{ \mid G \mid } \sum_{all g \in G}
 \sum_{h \in G \atop{I=odd}}\prod_{\tilde{\theta}_i^g=\tilde{\theta}_i^h=0}
 (1- {\frac{1}{q_i}})} \}-1
\end{equation}

\begin{equation}
\label{p21}
p_{2,1}= \{ {\frac{1}{2} \frac{1}{ \mid G \mid } \sum_{all g \in G}
\sum_{h \in G \atop{I=even}}\prod_{\tilde{\theta}_i^g=\tilde{\theta}_i^h=0}
(1- {\frac{1}{q_i}})} \}-1.
\end{equation}
(If the set of {\it i}'s satisfying $\tg=\th=0$ is empty, we define
$\prod_{\tilde{\theta}_i^g=\tilde{\theta}_i^h=0}(1- {\frac{1}{q_i}}) = 1$.)

It is a remarkable fact that we can calculate these numbers without any
information on actual structure of the chiral ring. These results are
natural
extentions of the formulae obtained by Roan\cite{ro}.He got simple formulae
for
 Betti numbers of resolved Calabi-Yau manifolds with $d = \sum{n_{i}}$
 in $WCP^4$, i.e. in the case of $N=5$ with $\sum{q_{i}} = 1$. Our formulae

 are applicable to other types of resolved Calabi-Yau manifolds,
 say Schimmrigk's construction of the Tian-Yau manifold and its orbifolds

 \cite{s,al} which can be described by $W(X_{i})$ with $N = 7$ and
$\sum{q_{i}} = 2$.

We can obtain the above results by the method of ref.\cite{ks3}, where
$P(1,1)$
 was used for calculation as well as $P(-1,-1)$. It is easy to compute
$P(1,1)$
 from our \p\ polynomial with projection operator and we get the same
result as
 in ref.\cite{ks3}.

\section{Conclusions and discussions}
In this paper we have constructed the \pp s with projection operators for
general
abelian \lgo s. Although we considered only three types of potentials, we
believe
that the same results hold for general $W(X_{i})$ \cite{sa}.
 If \cy\ interpretation exists, special dualities of $p_{i,j}$ are realized
  and simple formulae for $p_{1,1}$ and $p_{2,1}$ are obtained.

It is worth pointing out that our results will be useful to examie Yukawa
couplings. The geometrical method for Yukawa couplings are considered in
 \cite{c,cop} and it is a challenging problem to calculate them in terms of
  \lgo\ technique. Recent attempts are found in \cite{bh2}. To evaluate
  ${\bf {27^3}}$ Yukawa couplings, we must first get informations about
  (1,1) states. Using our formula for $p_{1,1}$ we can see which twisted
  sector coud have (1,1) states, without concrete representations of (1,1)

 states. Then we can apply the twist number selection rule to ${\bf 27^3}$
 Yukawa couplings and get non-vanishing ones.

{\it Acknowledgements} : The auther would like to thank Professor M. Ida
for
 helpful discussions and for careful reading of this manuscript.

\end{document}